\documentclass{eas}
\usepackage{amsmath}
\usepackage{graphicx}
\usepackage{natbib}
\usepackage{txfonts}
\usepackage{color}
\usepackage{bm}
\newcommand\bb[1] {   \mbox{\boldmath{$#1$}}  }
\newcommand\del{\bb{\nabla}}
\newcommand\bcdot{\bb{\cdot}}

\begin{document}

%
%
%


\def\jnl@style{\it}
\def\aaref@jnl#1{{\jnl@style#1}}

\def\aaref@jnl#1{{\jnl@style#1}}

\def\aj{\aaref@jnl{AJ}}                   
\def\araa{\aaref@jnl{ARA\&A}}             
\def\apj{\aaref@jnl{ApJ}}                 
\def\apjl{\aaref@jnl{ApJ}}                
\def\apjs{\aaref@jnl{ApJS}}               
\def\ao{\aaref@jnl{Appl.~Opt.}}           
\def\apss{\aaref@jnl{Ap\&SS}}             
\def\aap{\aaref@jnl{A\&A}}                
\def\aapr{\aaref@jnl{A\&A~Rev.}}          
\def\aaps{\aaref@jnl{A\&AS}}              
\def\azh{\aaref@jnl{AZh}}                 
\def\baas{\aaref@jnl{BAAS}}               
\def\jrasc{\aaref@jnl{JRASC}}             
\def\memras{\aaref@jnl{MmRAS}}            
\def\mnras{\aaref@jnl{MNRAS}}             
\def\pra{\aaref@jnl{Phys.~Rev.~A}}        
\def\prb{\aaref@jnl{Phys.~Rev.~B}}        
\def\prc{\aaref@jnl{Phys.~Rev.~C}}        
\def\prd{\aaref@jnl{Phys.~Rev.~D}}        
\def\pre{\aaref@jnl{Phys.~Rev.~E}}        
\def\prl{\aaref@jnl{Phys.~Rev.~Lett.}}    
\def\pasp{\aaref@jnl{PASP}}               
\def\pasj{\aaref@jnl{PASJ}}               
\def\qjras{\aaref@jnl{QJRAS}}             
\def\skytel{\aaref@jnl{S\&T}}             
\def\solphys{\aaref@jnl{Sol.~Phys.}}      
\def\sovast{\aaref@jnl{Soviet~Ast.}}      
\def\ssr{\aaref@jnl{Space~Sci.~Rev.}}     
\def\zap{\aaref@jnl{ZAp}}                 
\def\nat{\aaref@jnl{Nature}}              
\def\iaucirc{\aaref@jnl{IAU~Circ.}}       
\def\aplett{\aaref@jnl{Astrophys.~Lett.}} 
\def\apspr{\aaref@jnl{Astrophys.~Space~Phys.~Res.}}
\def\bain{\aaref@jnl{Bull.~Astron.~Inst.~Netherlands}} 
\def\fcp{\aaref@jnl{Fund.~Cosmic~Phys.}}  
\def\gca{\aaref@jnl{Geochim.~Cosmochim.~Acta}}   
\def\grl{\aaref@jnl{Geophys.~Res.~Lett.}} 
\def\jcp{\aaref@jnl{J.~Chem.~Phys.}}      
\def\jgr{\aaref@jnl{J.~Geophys.~Res.}}    
\def\jqsrt{\aaref@jnl{J.~Quant.~Spec.~Radiat.~Transf.}}
\def\memsai{\aaref@jnl{Mem.~Soc.~Astron.~Italiana}}
\def\nphysa{\aaref@jnl{Nucl.~Phys.~A}}   
\def\physrep{\aaref@jnl{Phys.~Rep.}}   
\def\physscr{\aaref@jnl{Phys.~Scr}}   
\def\planss{\aaref@jnl{Planet.~Space~Sci.}}   
\def\procspie{\aaref@jnl{Proc.~SPIE}}   

\let\astap=\aap
\let\apjlett=\apjl
\let\apjsupp=\apjs
\let\applopt=\ao

\title{Angular momentum transport in accretion disks: a hydrodynamical
  perspective}  
\runningtitle{Disks hydrodynamics}
\author{S\'ebastien Fromang}\address{Laboratoire AIM,
  CEA/DSM-CNRS-Universit\'e Paris 7, Irfu/Service d'Astrophysique,
  CEA-Saclay, 91191 Gif-sur-Yvette, France.} 
\author{Geoffroy Lesur}\address{Univ. Grenoble Alpes, CNRS, IPAG,
  F-38000 Grenoble, France.}
\begin{abstract}
The radial transport of angular momentum in accretion disk is a
fundamental process in the universe. It governs the dynamical
evolution of accretion disks and has implications for various
issues ranging from the formation of planets to the growth of
supermassive black holes. While the importance of magnetic fields for
this problem has long been demonstrated, the existence of a source of 
transport solely hydrodynamical in nature has proven more difficult to
establish and to quantify. In recent years, a combination of results
coming from experiments, theoretical work and numerical simulations
has dramatically improved our understanding of hydrodynamically
mediated angular momentum transport in accretion disk. Here, based on
these recent developments, we review the hydrodynamical processes that
might contribute to transporting angular momentum radially in
accretion disks and highlight the many questions that are still to be
answered.
\end{abstract}
\maketitle
\section{Introduction}

\noindent
One of the many research activities of Jean Paul Zahn during his
career was devoted to studying the mechanisms responsible
for outward angular momentum transport in accretion disks. More
specifically, he focused on the possibility that hydrodynamical turbulence
could develop and lead to enhanced and efficient transport: after all,
accretion disks are shear flows, and our experience of shear flows in
the laboratory and in our every day life is that they are easily
destabilized and prone to turbulence. As we shall see, this turned out
to be much more complicated than anticipated, to the point that the
very existence of such a hydrodynamic transition to turbulence in
accretion disks is still vividly debated today and far from being
settled. Nevertheless, Jean-Paul's approach to the problem
reinvigorated the interest of the astrophysical community for
hydrodynamic (as opposed to 
magneto--hydrodynamic) processes in accretion disks in general. This
renewed interest led to unexpected discoveries in the past few
years, and it is now clear that hydrodynamics processes are important 
if we want to reach a complete and coherent picture of accretion disks
dynamics. The aim of this review is to provide the community with a
description of these recent advances.

Before doing so, we start by recalling the problem at
hand. Accretion disks are objects of disk-like shape, mostly composed
of gas rotating around a central object of mass $M$. In this
review, we will restrict the discussion to situations for which
collisions between particles are frequent enough that the dynamics can
be described by the hydrodynamics equations, although we note that
this is not always the case, as for example in the inner parts of
disks rotating around supermassive black holes
\citep{quataert01}. Observations of accretion disks indicate that the
disk material falls on the central object. This is true for accretion
disks of all natures, regardless of whether the central object is a
supermassive black hole, a stellar mass compact object or a young star
\citep{fkr85}. It is precisely the existence of that accretion
flow that is difficult to understand theoretically 
because of angular momentum conservation: a particle with a given
angular momentum $\cal{L}$ with respect to the central object tends to
settle into a circular orbit around that object with a Keplerian
angular frequency $\Omega_K$ such that gravitational 
attraction is balanced by the outwardly directed centrifugal
force. For this particle to fall onto the central 
object, it must loose its angular momentum. Finding viable and
efficient mechanisms for doing so has been the subject of intense
research activities in the past $50$ years. Broadly speaking, one can
distinguish between two types of phenomena. Internal processes creates
an outwardly directed flux of angular momentum in the bulk of the
disk, for example in the form of turbulence or waves, wherease
external processes extract angular momentum from the disk by an
external torque. The later possibility encompasses disks winds and
jets that might be associated with a large scale magnetic field for
example \citep{FRC14}. In this review, we will focus on the former,
and only discuss occasionally the connection that might exist between
both. 

The processes operating within the disk can themselves also be
separated in two classes, depending on whether or not they depend on
the presence of a magnetic field. This review is devoted to processes
that operate in disk in the absence of a magnetic field. However, the
field has been dominated in the past thirty years by research
connected to the so--called magneto-rotational instability \citep[MRI,
][]{balbus&hawley91,balbus&hawley98}, to the point that many readers may
wonder whether it is still even worth considering hydrodynamic
processes at all. For this reason, we start in section~\ref{sec:mri}
by discussing the difficulties that remain associated with the MRI
paradigm, but also take the opportunity to quickly review the latest
results in that field, some of which are relevant for the present
review. In section~\ref{sec:hydro_stab}, we go back to the pure
hydrodynamic case and briefly discuss the linear stability of
accretion disks in this context. We next
focus in section~\ref{sec:subcrit_turb} on the possible existence of a
subcritical transition to turbulence in these objects, highlighting
the contribution of Jean-Paul Zahn in this field. As we shall see, the
question is not yet settled, and several researchers have gone to
explore other routes in the past few years. In particular, much
attention has been devoted, with some success, to the consequences of
the peculiar thermodynamics of protoplanetary disks. We thus describe
the most promising of these results in 
section~\ref{sec:thermo_effect}. Finally, we briefly discuss in
section~\ref{sec:misc_proc} some of the many other possibilities that
have been proposed or are currently being debated in the literature
to account for angular momentum transport in accretion disks.

\section{The magnetorotational instability and its shortcomings}
\label{sec:mri}

\noindent
The question of the potential relevance of hydrodynamics effects in
disks may be deemed irrelevant because we know there exists a powerful 
magnetohydrodynamical (MHD) instability in accretion disks: indeed,
flows in keplerian rotation around a central object are destabilized 
by a weak magnetic field. Although known since the 1960's
\citep{velikhov59,chandra61}, the relevance of that instability for
accretion disks was only realized three decades later by
\citet{balbus&hawley91}. The discovery of the magneto--rotational
instability (MRI), as it was to be called, revolutionized our
understanding of angular momentum transport in accretion disks. This
is because the MRI grows on dynamical timescales, has minimal
geometrical requirements (a subthermal B-field and a radially
decreasing angular velocity) and leads during its nonlinear evolution
to fully developed MHD turbulence and an associated outward flux of
angular momentum that is roughly (i.e. to within an order of
magnitude) compatible with observational constraints
\citep{balbus&hawley98}. At first glance, it appears that the MRI is
the ideal solution to angular momentum transport in accretion disks.

However, the MRI paradigm is not without problems. Indeed, in the cold
and dense protoplanetary (PP) disks, the ionization fraction is so small
that the gas and the field are not well coupled: ideal MHD does not
apply and non-ideal effects cannot be ignored. Considering the
stabilizing effects of ohmic resistivity on the linear instability,
\citet{gammie96} first showed that many regions of PP
disks are immune to the MRI: accretion proceeds only through the disk
upper layers thanks to MRI--driven MHD turbulence, while their
midplane remains laminar. This layered accretion scenario was
later confirmed using numerical simulations \citep{fleming&stone03} 
and the dead zone paradigm became the dominant and widely
accepted scenario during the following decade. However, it only
includes the effect of Ohmic diffusion, whereas typical
densities and temperatures expected in PP disks show that
both ambipolar diffusion and the Hall effect should be dominant in
most parts of these objects. Both processes turned out to affect the
conclusions of \citet{gammie96}. The addition 
of ambipolar diffusion in the layered accretion paradigm leads to
an even more dramatic picture where MRI turbulence is severely
suppressed in most of the disk, leading to accretion rates at least 10
times lower than observed ones \citep[see][ for a
  review]{turneretal14}. In order to ``save the MRI'', magnetised disk
winds, mostly neglected since the discovery of the MRI, came back to
the scene as a viable alternative to MRI turbulence. Effectively, the
MRI naturally produces magnetised outflows upon saturation
provided that the disk is threaded by an external poloidal field
\citep{suzuki&inutsuka09,suzukietal10,bai&stone13,fromangetal13}. This
wind, launched from the very surface of the disk ionised by far-UV
radiations, could be sufficient to recover accretion rates compatible
with observations \citep{bai16}. The 
addition of the Hall effect leads to an even more complicated picture
since it is a purely dispersive effect, which does not lead to energy
dissipation. It induces significant modifications to the flow
topology, including the appearance of strong 
zonal flows \citep{kunz&lesur13} and/or an increased importance of
disk winds in torquing the disk \citep{lesuretal14,bai14,bai15}. Many
aspects of Hall mediated accretion remain to be elucidated, largely
because these results have been obtained in a local framework and may
suffer from undesired but uncontrolled artifacts. Nevertheless, at the
time this review is being written, we cannot rule out the possibility
that angular momentum transport in PP disks is regulated
by processes at least partially governed by Hall physics.

Given all these uncertainties, it is very tempting to consider
alternative routes to angular momentum transport that do not rely on
MHD processes. This motivated several groups, including Jean-Paul
Zahn's, to investigate purely hydrodynamical instabilities.

\section{The hydrodynamical stability of accretion disks}
\label{sec:hydro_stab}

In cylindrical coordinates $(R,\phi,z)$, the stability of
non-magnetised rotating sheared flows is well 
described by the Solberg-H\o iland criterion, which is a
\emph{sufficient} condition for \emph{stability} under the action of
\emph{infinitesimal axisymmetric and adiabatic} perturbations
\citep{tassoul07}: 
\begin{align}
\label{eq:SH1}\frac{1}{R^3}\frac{\partial j^2}{\partial R}
-\frac{1}{C_p\rho}\bm{\nabla}P\cdot\bm{\nabla S}&>0 
\end{align}
and
\begin{align}
\label{eq:SH2}\frac{\partial P}{\partial z}\left(\frac{\partial
  j^2}{\partial R}\frac{\partial S}{\partial z}-\frac{\partial
  j^2}{\partial z}\frac{\partial S}{\partial R}\right)&<0, 
\end{align}
where $j$ is the specific angular momentum, $C_p$ the calorific
capacity at constant pressure, $\rho$ the density, $P$ the pressure
and $S$ the entropy of the gas under consideration. 

To illustrate how this criterion applies to accretion disks, let us
consider a simple accretion disk model with a Keplerian rotation
profile 
\begin{equation}
\Omega_K=\sqrt{\frac{GM}{R^3}}.
\end{equation}
The first Solberg-Ho\"iland criterion (\ref{eq:SH1}) can be recast as
\begin{align}
\label{eq:SH1_reduced}\Omega_K^2+N_R^2>0	
\end{align}
where we have introduced the radial Brunt-V\"aiss\"al\"a frequency $N_R$
\begin{equation}
N_R^2=-\frac{1}{\gamma \rho}\frac{\partial P}{\partial R}
\frac{\partial}{\partial R} \ln \left( \frac{P}{\rho^\gamma} \right)
\, .
\end{equation}
 In thin disks, where the disk thickness $H$ is much smaller than the
 disk radius $R$, we have $|N_R^2|\sim (H/R)^2$, so that for all
 practical purposes, this criterion is always satisfied. In the case
 where radial buoyancy vanishes (and hence entropy is constant over
 radius), we recover the usual Rayleigh criterion which predicts
 stability for flows with increasing specific angular momentum $j$.  

In this simple disk model, the rotation profile $\Omega_K$ only
depends on $R$. For this reason, the second Solberg-Ho\"iland
criterion (\ref{eq:SH2}) reduces to the Schwarzschild criterion for
vertical convection: 
\begin{align}
N_z^2>0,	
\end{align}
with the vertical Brunt-V\"aiss\"al\"a frequency $N_z^2$. This
criterion is usually satisfied in PP disks where the disk
surface is hotter than the disk midplane due to external
irradiation. However, some disks heated by turbulent dissipation
sometime violate this criterion, especially when the gas reaches very
high opacities, such as in dwarf novae, and convection might develop.
(see sec.~\ref{sec:misc_proc}).

\section{Subcritical transition to turbulence}
\label{sec:subcrit_turb}

\subsection{The basic idea}

\noindent
As shown above, the radial profile of angular velocity in accretion
disks is Keplerian and is therefore stable to axisymmetric
infinitesimal perturbations. However, linear axisymmetric stability is
not necessarily the full story. There are many examples of
linearly stable flows that become unstable and develop turbulence at
large Reynolds numbers such as Poiseuille (pipe) flows and plane
Couette flows. Such a transition requires finite amplitude (as opposed
to vanishingly small) fluctuations and is called a subcritical
transition. While linearly stable at all Reynolds number $Re$,
turbulence is easily observed in plane Couette experiments whenever
$Re$ reaches $\sim 1600$ \citep{ta92}. The later is typically defined
as 
\begin{equation}
Re=\frac{SL^2}{\nu} \, ,
\end{equation}
where $\nu$ stands for the fluid kinematic viscosity, while $S$ is the
shear rate and $L$ is the distance between the walls. In accretion
disks, order of magnitude estimates of $Re$ are fairly easy to
obtain. For example, \citet{fkr85} quote $Re \sim 10^{14}$ as being
typical for disks around white dwarfs in binary systems. This is a
general result: $Re$ is always larger by many orders of magnitudes
than the typical critical Reynolds numbers above which turbulence is
found for most linearly stable shear flows in the laboratory. These
simple arguments led many researchers to conjecture that a nonlinear
transition to turbulence might also exist in accretion disks and be
responsible for outward angular momentum transport in these objects.

\subsection{Jean Paul Zahn's era}

\noindent
Finding such a transition, however, proved more challenging than one
might naively think. This is because of the Coriolis force. Compared
to standard Couette flows, it promotes the occurrence of epicyclic
motions and has a stabilizing influence on the flow
\citep{balbusetal96}. In fact, early numerical simulations, performed
in the framework of the so--called shearing box
\citep{goldreich&lyndenbell65}, a well-known 
local model of accretion disk, failed to find any signature of
turbulence in Keplerian flows
\citep{balbusetal96,hawleyetal99}. However, the limited spatial
resolution that could be reached at that time, associated with the
uncontrolled dissipation of the numerical method that was used,
prevented definite conclusions to be drawn from these results. 

This is when and why Jean Paul Zahn started to work on the
problem. Along with his collaborators and PhD students, he realized
that fluid experiments could help improve our understanding of the
problem. In this regard, the Taylor-Couette flow is particularly
interesting: it consists in two concentric
rotating cylinders sandwiching a fluid, which is generally an
incompressible liquid. Drag exerted by the rotating cylinders onto the 
liquid sets the later in rotation. By varying the angular velocities
of both cylinders, the angular velocity and angular
momentum radial profile of the flow within the cavity and its Reynolds
number can be varied. While the Taylor-Couette apparatus has been the
subject of active research since the pioneering work of
\citet{taylor1923}, the regime closest to accretion disk had only been
barely considered by the end of the 20$^{\textrm{th}}$ century. 

The contribution of Jean-Paul Zahn was twofold: first, along with his
students, he re-analyzed results from old experiments performed in a
regime of radially increasing angular momentum by Wendt (1933) and
Taylor (1936). These old results hinted toward a transition to a
chaotic flow, possibly turbulence, at large Reynolds number
\citep{richard&zahn99}. But Jean-Paul Zahn also fully realized the 
shortcomings of these conclusions, based on a series of old
experiments. This is why he initiated a collaboration with a team of
physicists at the CEA with whom he embarked into building his own
Taylor-Couette experiment (see figure~\ref{fig:TC}, left panel). Using
this new device, the group studied, 
for the first time with modern techniques, the regime most analogous
to accretion disks. While claiming, by visual inspection, a transition
to turbulence, they also failed to properly characterized its
properties \citep{richard01}. In their conclusions, they noted that an
improvement of the treatment of the ends caps of the experiment as well
as extending the investigation toward higher Reynolds numbers (their
largest $Re$ value is of the order of a few times $10^4$) were both
needed. But the stage was set for future investigation. And in fact,
\citet{richard&zahn99}  close their paper with the following sentence:
{\it ``(...) we hope that experimentalists will turn again to
this classical problem, which is of such great interest for
geophysical and astrophysical fluid dynamics, and that they will
explore the rotation regimes for which the data are so incomplete.''}
As we shall see now, their hope has been largely fulfilled...

\subsection{The current state of affair}

\begin{figure}
\begin{center}
\includegraphics[scale=0.4]{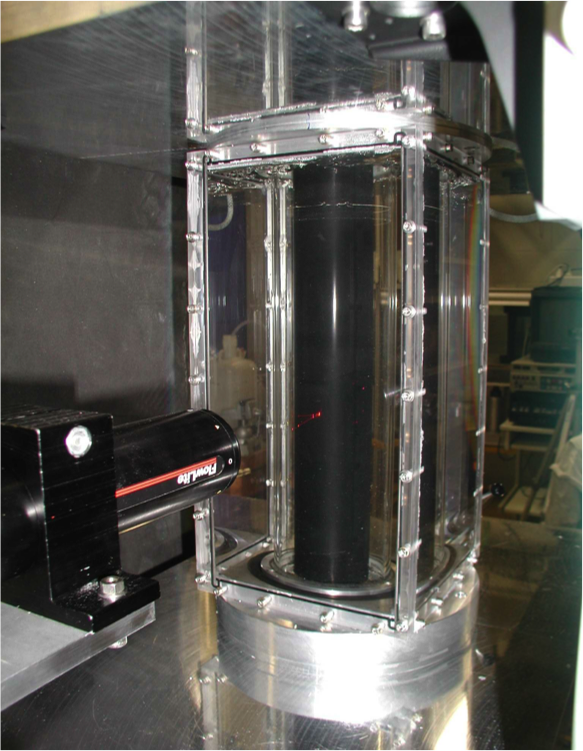}
\includegraphics[scale=0.61]{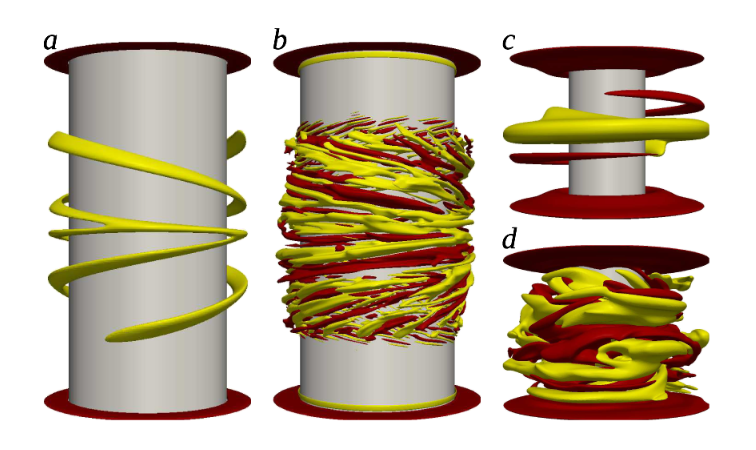}
\caption{Left: the CEA Taylor-Couette experiment built in
  collaboration with Jean-Paul Zahn in the early 2000's (Credit:
  CEA/DSM/IRAMIS/SPEC). Right: 
  Isosurface of the radial velocity in simulations of the Maryland
  (panels a and b) and Princeton (panels c and d) experiments at
  $Re=1332$ (a), $Re=5328$ (b), $Re=1545$ (c) and $Re=6437$ (d). From
  \citet{avila12}.} 
\label{fig:TC}
\end{center}
\end{figure}

\noindent
Jean Paul's efforts to motivate the scientific community toward
experimental astrophysics has been successful indeed: the regime of
outwardly increasing angular momentum flows is now the subject of
active research by the community of experimentalists experts in
Taylor-Couette flows. In that respect, the interested reader will
benefit from the recent review of \citet{grossmannetal16}.

What did we learn from this intense research activity? The least we
can say is that the status of the field is confused! Some groups
report that Keplerian flows become turbulent, some other that they
remain laminar. The experiments that paid particular attention to the
regime most relevant for accretion disks can be divided in two
groups. On the one hand, the Twente \citep{vangilsetal11} and Maryland
\citep{paoletti&lathrop11} experiments both report that the flow is
turbulent. The claim is based on measurements of an increased torque
exerted by the fluid on the rotating cylinders compared to what would
be expected from a laminar flow. The results are consistent
between the two experiments \citep{paolettietal12}. On the other hand,
the Princeton experiment \citep{jietal06,schartmanetal12,edlund&ji14}
reports that the flow remains laminar up to a Reynolds number equal to
$2 \times 10^6$, a value comparable in amplitude to that
reached by both the Twente and Maryland experiments. The claim is
based on doppler velocity measurements. In addition to confirming
this result, \citet{schartmanetal12} measured increased velocity
fluctuations near 
the end caps of the Princeton apparatus (see their figure 4 and 7),
which they suggest might arise because of an unstable shear layer that
develops at that location. To our knowledge, there is no real
understanding accepted in the community that would explain these
conflicting results. The experiment devices have differences: the
Twente and Maryland apparatus are tall and narrow (the height $L$
ranges between $70$ and $90$ centimeters and the width $\Delta R$ is
of the order of $6$ cm), 
while the Princeton experiment is small and wide ($L=28$ cm and
$\Delta R \sim 13$ cm). Such a difference in the apparatus
``curvature'' (quantified as the ratio $\Delta R/R$) is known to
significantly affect the critical Reynolds number of subcritical
transitions \citep{richard&zahn99}, making turbulence more difficult
to reach in a wide apparatus. Using the scaling proposed by
\cite{richard&zahn99}, the critical Reynolds number of the Princeton
experiment should be around $6\times 10^5$ \citep{jietal06}, 
i.e. very close to the maximum Reynolds number reached in this
experiment (a few times $10^6$). Therefore, there is a possibility
that the Princeton experiment is below or right at the 
critical Reynolds number of the transition due to the dimensions of
the apparatus, leading to a laminar flow. Another difference lies in
the treatment of the end caps of the experiments: both axial
boundaries corotates with the outer cylinder in the Twente and
Maryland experiments, while they are 
composed of two independently rotating rings in the Princeton
experiments. In order to investigate how these differences might
affect the flow, \citet{avila12} performed dedicated direct numerical
simulations of the two experiments. Of course, being limited by
current computational capabilities, the maximum Reynolds number that
could be reached in the simulations, of order a few thousands, is much
smaller than that of the experiments. Nevertheless, the results show
that the flow becomes turbulent in both simulations (see
figure~\ref{fig:TC}, right panel) and that this
transition is due to the boundaries. This has been confirmed by
recent simulations at Reynolds number up to $\sim 50 000$ where 
turbulence was found to be more and more confined to the vicinity of
the end caps as $Re$ increases, while the bulk of the flow remains
laminar \citep{lopez&avila17}: finite size effects are definitely
important in the experiments and should be investigated carefully. 

In order to free themselves from such complications associated with
the boundaries, but also to consider geometries that would be
closer to actual accretion disks than Taylor-Couette experiments, a
few groups performed numerical simulations that addressed the question
of the subcritical transition to turbulence using different numerical
setups. First, \citet{lesur&longaretti05} revisited the results
presented by \citet{balbusetal96} and \citet{hawleyetal99}. Performing
direct numerical simulations in the framework of the shearing box,
they tried to evaluate the evolution of the 
critical Reynold number $Re_c$ (above which the flow becomes turbulent) as
one enters the Rayleigh stable regime. They concluded that $Re_c$ lies
in the range $10^{10-26}$ for Keplerian flows and that such large
critical Reynolds number 
would lead to angular momentum transport rates that would be, for all
astrophysical purposes, much smaller than required by the
observations. The weakness of their approach, though, lie in the
massive extrapolation involved because only states very close to the
Rayleigh stable line could be investigated numerically due to the
massive computational requirements required when $Re$ grows above a
few thousands. This 
explains the enormous range in the value of $Re_c$ quoted above. 
More recently, and benefiting from the modern computational resources
that are now available, \citet{ostillamonicoetal14} investigated
Taylor-Couette flows with periodic axial boundaries (so that they
avoid the complexity associated with the end caps discussed above) for
Reynolds numbers up to $\sim 10^5$. Their strategy consisted in
starting from a turbulent flow (obtained in the Rayleigh unstable
regime by choosing an outer cylinder at rest), and suddenly enter the
stable regime by setting the outer cylinder in rotation. In all their
simulations, \citet{ostillamonicoetal14} found that turbulence decays,
although with longer and longer timescales as $Re$ goes up. Both of
the aforementioned studies agree in the fact that $Re_c$ is larger
than $10^5$. 

Because of the finite computational ressources at our disposal and the
gigantic $Re$ of accretion disks, numerical simulations will always
suffer from the limitations hightlighted above. Alternative approaches
are thus desirable. This is the purpose 
of methods based on the so-called linear transient growth approach,
such as originally proposed for this problem, in collaboration with
Jean Paul Zahn, by \citet{TC03},  and further described in this volume
by Gogichaishvili et al. (2017). As is well known, the linear operator
describing rotating sheared  flows is non-normal, so that linear
perturbations can see 
their energy grow by several orders of magnitude for a limited amount
of time before being ultimately dissipated. It has
been suggested that these transient amplifications could be the route
to turbulence in sheared flows since transiently amplified modes could
become non-linear and excite new modes which could be amplified again,
etc...The key question here is to know whether non-linear interactions
can regenerate sufficiently rapidly new transient modes. It is
therefore critical to understand how this non-linear feedback proceeds
(or fails) in astrophysical flows, a point that still deserves to be
established. We end by stressing that, while having
transient amplification might be required for the flow to be subject
to a subcritical transition to turbulence, it is clearly \emph{not a
sufficient condition}, as recently illustrated by
  \citet{shietal17}. For all these reasons, the importance of such 
transiently growing disturbances for angular momentum transport in
accretion disks still remains to be firmly established. An
alternative and maybe more promising approach is to focus on isolating
steady and/or oscillating structures in the flow that would form the
skeleton for the turbulent dynamics. Based on this idea,
\citet{ROC07} used continuation methods to find steady nonlinear
solutions for Rayleigh stable cyclonic flows (i.e. $d\Omega/dR>0$) but
failed to do so for anticyclonic flows (and therefore Keplerian
profiles), for which $d\Omega/dR<0$, suggesting the later were more
difficult to destabilise by non-linear instabilities than the
former. Although not fully conclusive, such an approach could serve as
a starting point for future theoretical work.

More than $20$ years after Jean-Paul Zahn first tried to address this 
problem, the question of whether or not accretion disks are vulnerable
to a subcritical transition to turbulence cannot yet be unambiguously
answered. 

\section{The importance of the disk thermodynamics}
\label{sec:thermo_effect}

\noindent
In real life, accretion disks are not composed of an incompressible 
liquid rotating between rigid cylinders! Some of them are irradiated
by the central object around which they rotate, while all of them
radiate away some of their internal energy into space. In other words,
their thermal structure is governed by complex processes. In recent
years, this has been shown to be important for their stability. The
purpose of the present section is to discuss these recent results,
which have mostly been obtained in the context of PP disks. In order
to fix ideas, we start by recalling the properties of these objects
and their peculiar thermal structure. 

\subsection{Protoplanetary disk properties and thermal structure}
\label{sec:disk_thermal_struct}

\noindent
PP disks are found rotating around newly born stars. Thanks to the
wealth of observations that are now available, we have a fairly good
idea of their basic properties \citep[see][ for a
review]{williams&cieza11}. They 
are mostly composed of gas, but also contain about $1 \%$ in mass of solid
particles (usually refereed to as dust), which is very important for
setting disks opacities. Incidentally, dust particles are also
believed to be the building blocks of planets. Disk masses are very
uncertain because they are mostly composed of optically thick hydrogen
which is actually very difficult to see, but the estimates are of the
order of a few percent of the central stellar mass. Their lifetime,
also uncertain, is of the order of a few millions years.

To first order, PP disks are essentially passive disks: their thermal
structure is set by a balance between heating due to the irradiation
of the disk upper layers by the central star, mostly at UV and visible
wavelengths, and cooling due to the dust particles thermally radiating
their energy to space at infrared wavelengths
\citep{chiang&goldreich97}. Simple 1+1D models can be constructed
and enable the $(R,Z)$ temperature distribution to be estimated with a
fair degree of accuracy \citep[see][ for a
  review]{dullemondetal07}. They show that the 
midplane temperature of PP disks decays with the distance
$R$ to the central star (with a power law scaling close to
$R^{-1/2}$). Its vertical profile is essentially isothermal outward of
approximately a couple of astronomical units (AUs) in the bulk of the
disks except for the very tenuous surface layers where it increases
due to stellar irradiation. Internal
heating, such as for example induced by turbulent dissipation, only
starts playing a role in the few inner tens of AU
\citep{dalessioetal98}. 

In recent years, these simple models have been extended to include
more realistic treatments of the radiative transfer 
(opacity, geometry) and additional physical processes (dust and
gas decoupling, chemistry). Perhaps the
most realistic disks structure that have been published so far are the
radiation chemico-dynamical models that can be calculated using the
code PRODIMO \citep[see][ and the subsequent improvements of the
method]{woitkeetal09}. They include dust continuum radiative transfer,
gas-phase and photo-chemistry, as well as gas thermal balance, and
compute the disks hydrostatic vertical structure assuming it is
axisymmetric. Detailed calculations using PRODIMO roughly confirm the
disk structure described above and can be summarized as
follows: radially decreasing and relatively cold temperature in the
midplane, sandwiched between hot surface layers where dust and gas
temperatures decouple.

Disk temperatures, however, are not neccessarily constant and may vary
with time: how important an effect it is depends on how fast the
material can heat and cool. This effect can be evaluated by means of
the cooling time  $\tau_c$. In the midplane, cooling is mediated by
dust particles and 
is efficient, so that $\tau_c$ is only a fraction of the orbital
timescale \citep{woitkeetal09,lin&youdin15,kratter&lodato16}. $\tau_c$
increases in the low-density upper layers of the disks where it is
dominated by atomic lines cooling and becomes longer than the orbital
timescale. For example, at $10$ AU, PRODIMO calculations suggest that
$\tau_c \sim 50$ to $100$ years at locations a few scaleheights above
the disk midplane \citep[][ see their figure 13]{woitkeetal09}.

\subsection{The baroclinic instability (SBI) and the convective over-stability}

\noindent
The ideas developed above lead to the realization that the radial
thermal structure of PP disks may play a role in their
dynamics. However, this route to instability is not straightforward 
since the linear stability criterion 
(\ref{eq:SH1_reduced}) shows that thin disks are linearly stable to
adiabatic perturbation. The key here it to realise that perturbations
are not necessarily adiabatic in PP disks, essentially
because the cooling timescale can be of the order of the orbital
timescale in the midplane (see discussion above). This fact alone is
enough to violate the Solberg-Ho\"iland 
criterion, but it is not enough to make the flow unstable! Indeed, the
baroclinic instability and the convective over-stability are both
thermal instabilities. Their source of free energy is not the shear
but the thermal structure of the flow. For this reason, in addition to
non-adiabatic perturbations, they also require the Schwarzschild
criterion for radial convection to be satisfied, i.e. $N_R^2<0$. 

\subsubsection{The baroclinic instability (SBI)}

\begin{figure}
\begin{center}
\includegraphics[scale=0.4]{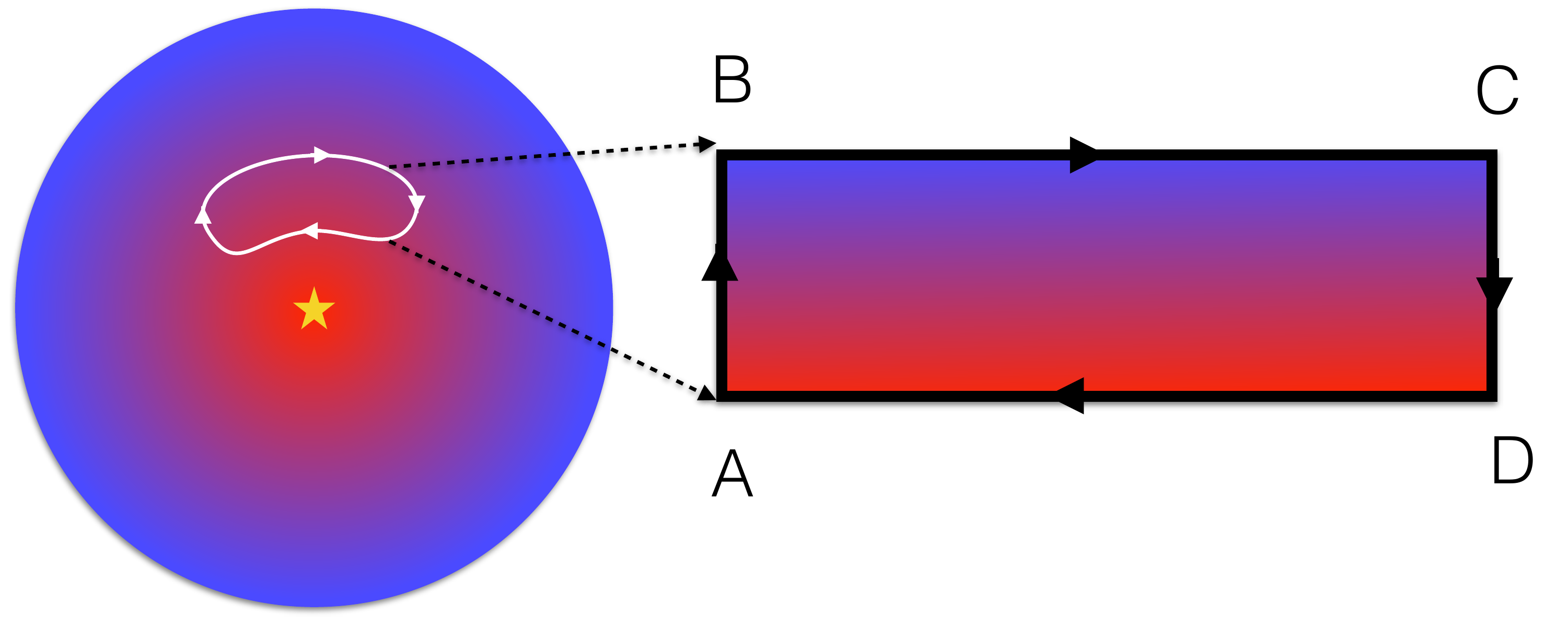}
\caption{Schematic illustration of the subcritical baroclinic
  instability. During its path along a vortex stremline (left panel),
  idealized as a rectangle (right panel), a fluid particle is
  buoyantly accelerated on the radial branches (A-B and C-D) and
  thermalized on the azimuthal branches (B-C and D-A). See text for
  details. Adapted from \citet{lesur&pap10}.}
\label{fig:SBI}
\end{center}
\end{figure}

\noindent
The baroclinic instability arises in 2D $(R,\phi)$ razor-thin disks in
the presence of non-axisymmetric disturbances when the two conditions
above are satisfied ($N_R^2<0$, finite $\tau_c$). Early simulations by 
\citet{klahr&bodenheimer03}, later supported by numerical evidences 
from \citet{petersenetal07a,petersenetal07b}, first reported the
spontaneous growth and long term survival of vortices in this situation. 
It even appeared that the associated density
fluctuations, driven by the coupling between incompressible vortices
and sound waves \citep{heinemann&pap09}, were able to create a
significant angular momentum flux 
\citep{klahr&bodenheimer03}. However, some of these results were
sensitive to the numerical properties of the algorithm
\citep{johnson&gammie05} and the linear stability analysis, suggesting
only transient amplification of linear disturbances
\citep{klahr04,johnson&gammie05}, was difficult to interpret, so that
the very existence of a baroclinic instability in PP
disks was subject to caution. Later numerical results, however, helped
clarify its existence, origin and properties: The baroclinic
instability exists but it is, in fact, subcritical: finite 
amplitudes disturbances are necessary to reach a turbulent
state \citep{lesur&pap10}. For this reason, the baroclinic instability
is now refereed to as the subcritical baroclinic instability, or SBI
for short. The subcritical nature of the instability explains the
failures of the linear analysis mentioned above to unambiguously
identify the mechanism at work.

Based on these numerical evidences, \citet{lesur&pap10} gave a simple
physical interpretation to explain the growth of vortices induces by
the SBI. Within an emerging vortex, which can be idealized as a
rectangle such as shown on figure~\ref{fig:SBI}, fluid particles are buoyantly
accelerated on the two radial branches (thus the requirement $N_R^2<0$)
and exchange heat on both horizontal branches (thus the requirement
for a finite heat diffusion) so that they are thermalized when
starting the next radial branch where they will be accelerated again,
allowing the vortex to strengthen in the process.

In the past few years, these results have been independently confirmed
\citep{lyra&klahr11,raettigetal13} and extended to more realistic
geometries that include the disk density vertical stratification
\citep{bargeetal16}. Our knowledge of the SBI can be summarized as
follows. First, it is severely suppressed in the presence of a
magnetic field \citep{lyra&klahr11}, so that it can only affect those
regions of the disk known as the dead zone (see
section~\ref{sec:mri}). The  saturation of the SBI occurs through two 
mechanisms: the first involve 
centrifugal and parametric instabilities that destabilize the vortices
themselves and prevent an unbounded growth \citep{lesur&pap09}. Being
of small scale, they tend to be difficult to capture in numerical
simulations. The second is through the emission of sound waves,
modified by rotation, that take the form of spiral waves propagating
radially away from the vortices. These spirals also have two
consequences. First, they create an outward flux of angular momentum
in the disk. It is customary in the literature to measure
that flux using a dimensionless parameter called $\alpha$\footnote{There
are several definitions of $\alpha$ in the literature, all of which
are consistent to within a factor of a few. Observational constraints,
based on typical evolutionary timescales of the various objects,
suggest that $\alpha$ values typically range between $10^{-3}$
and a few times $10^{-1}$}. In this particular case, $\alpha \sim $ a
few times $10^{-3}$ \citep{lesur&pap10,raettigetal13}. Second, they
induce a systematic and fast inward migration of vortices
\citep{paardekooperetal10}, with can be considered to be a negative
feedback on the saturation of the SBI. All these results, however,
were established using very idealized simulations, so that the
astrophysical implications of the SBI still remain to be investigated
and quantified in more details.

\subsubsection{The convective over-stability}

\noindent
The convective over-stability is an other flavour of the same process
which has been discovered in axisymmetric 3D disks
\citep{klahr&hubbard14}. This instability is linear, and its growth
rate can be estimated as
\begin{equation}
\gamma\simeq -\Omega
\Bigg(\frac{N_R}{\Omega}\Bigg)^2\frac{\Omega\tau_c}{1+\Omega^2\tau_c^2}, 
\end{equation}
from which we recover easily the two conditions for instability
discussed above. The optimum growth rate is found when $\tau_c= 1$ and
assuming the disk is thin, it scales like $\Omega (H/R)^2$ which
implies a relatively low growth rate in astrophysical applications. 

Physically, it relies on the same phenomenology as the one discussed
for the baroclinic instability (fig.~\ref{fig:SBI}), except that fluid
particles are not following a vortex streamline but are instead
undergoing an epicyclic orbit. At each oscillation, fluid particles
undergo a radial acceleration due to buoyancy, and cooling allows the
flow to thermalise between two radial displacements. Averaged over one
oscillation, this feedback loop leads to an amplification of epicyclic
oscillations, hence the name of the instability \citep{latter16}. 

Overall this instability produces global horizontal oscillations,
which if sufficiently amplified could break up into
turbulence. Therefore, it has been suggested that the convective
over-stability could act as a seed for the subcritical baroclinic
instability \citep{lyra14}. How this conclusion depends on $\tau_c$
and $N_R/\Omega$ however remains to be clarified. 

\subsection{The vertical shear instability (VSI)}
\label{sec:vsi}

\begin{figure}
\begin{center}
\includegraphics[scale=0.18]{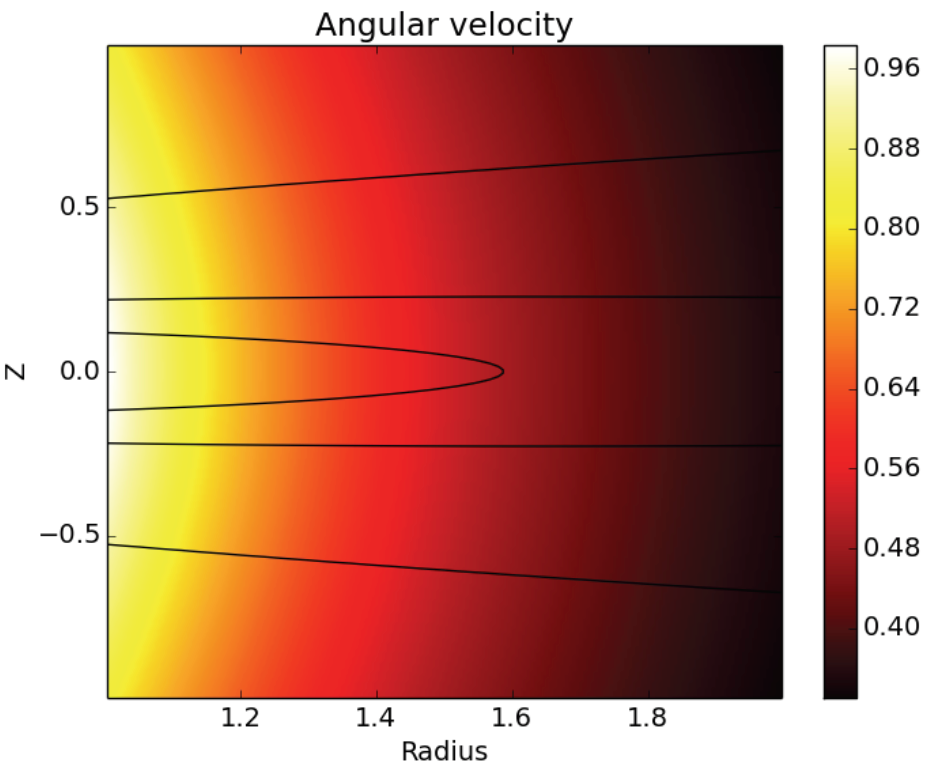}
\includegraphics[scale=0.18]{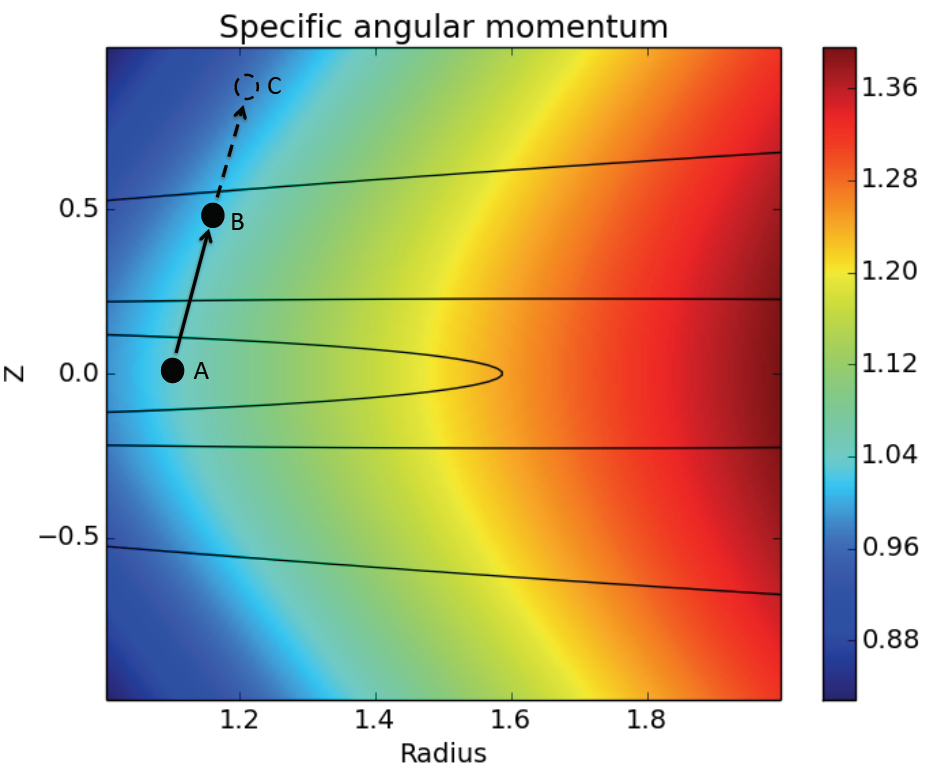}
\caption{Cartoon illustrating the physical mechanism of the
  Vertical Shear Instability (VSI). Angular velocity (resp. specific
  angular momentum) contours are shown on the left (resp. right) panel
  for a locally isothermal disk as a function of $R$ and $z$. On both
  panels, the disk
  density is shown using the solid contours, with levels equals to
  $0.5$, $0.1$ and $10^{-5}$ times the disk maximum density. A fluid
  particle, initially located in $A$, finds itself in $B$ with an
  excess angular momentum after a displacement along the black line
  (right panel)
  and keep moving upward and outward along the dashed line to $C$. See text
  for details.
}
\label{fig:vsi_linear}
\end{center}
\end{figure}

\noindent
Another consequence of PP disks thermal structure is the vertical
shear instability (VSI), which is the disk 
analog of the Goldreich-Schubert-Fricke instability that has been
familiar to the specialists of stellar dynamics since the 1960's
\citep{goldreich&schubert67}. By contrast, its potential relevance for
accretion disks dynamics was only mentioned much later in the
literature by \citet{urpin&brandenburg98} and \citet{urpin03}, 
and the first numerical evidences for its existence have only been
reported recently by \citet{nelsonetal13} using global numerical
simulations of locally isothermal disks (in which case the temperature
$T$ depends only on the cylindrical radius $R$). As explained by
\citet{barker&latter15}, the VSI is fundamentally a centrifugal
instability that relies on the existence of a vertical gradient of
$\Omega$, a well known consequence of baroclinicity in accretion
disks. By taking the curl of the force balance relation, it is indeed
easy to show that the baroclinic terms relates to the vertical
gradient of $\Omega$ through:
\begin{equation}
\frac{\partial R\Omega^2}{\partial R}=-\frac{1}{\rho^2} \left( \del
\rho \times \del P \right) \bcdot \, \bb{e_{\varphi}} \, .
\end{equation}
The vertical gradient of $\Omega$ (see also
figure~\ref{fig:vsi_linear}, left panel) implies that the second
Solberg-H\"oiland criterion (\ref{eq:SH2}) now includes the term
$\partial j/\partial z$, and this opens up the possibility of 
violating this criterion. Following \citet{barker&latter15}, we
illustrate such a possibility for the idealized and simple case
of a locally isothermal disk and for vanishingly small cooling times
$\tau_c$ for which buoyancy forces vanish. The contours of angular
velocity and specific angular momentum are shown on
figure~\ref{fig:vsi_linear} for this model, respectively on the left
and right hand side panels: consider the fluid particle initially in
A, with specific angular momentum ${\cal 
  L}_A$. If it is displaced along the black line, it will find itself
in position B with an excess angular momentum (because ${\cal
  L}_A>{\cal L}_B$), in clear violation of the Rayleigh criterion, and
will keep moving upward and outward.

In real disks, however, the vertical entropy profiles
tends to be stabilising. Buoyancy forces oppose the motions just
described, and some amount of thermal diffusion (or radiative cooling)
is necessary to prevent a complete stabilization of the VSI (which
materializes in an adiabatic, or isentropic, disk by the vertical
gradient of $\Omega$ vanishing). Once again, the disk thermal
properties turn out to be a key ingredient of the process! In
addition, viscosity also stabilizes the smallest unstable scales
\citep[as shown early on by ][]{nelsonetal13}. For these reasons, the
VSI is in fact a double-diffusive instability \citep{barker&latter15}:
it develops at 
intermediate scales for which neither viscosity nor buoyancy dominate
the fluid motions. \citet{lin&youdin15} have recently coupled these
simple arguments with analytical models of PP disks
to show that the VSI should be expected to develop at radii between a
few AUs and $\sim 100$ AUs thanks to sufficiently short cooling times
at those locations.

The simple physical interpretation of the VSI discussed above should
not hide the complexity of the linear modes of the instability, which
have been the subject of detailed calculations in the recent literature 
\citep{nelsonetal13,barker&latter15,umurhanetal16a}. These modes 
come in two flavors: the first type is composed of inertial waves that
are destabilized by the vertical angular velocity gradient. They have
modest growth rates and grow at relatively long wavelengths. The second
type is composed of surface modes that grow faster and at smaller
wavelengths set by the flow kinematic viscosity. For these reasons,
they are more difficult to capture in numerical simulations, and are
likely to be affected by numerical effects in the simulations that
have been published so far. For this reason, it is still presently
unclear which class of modes will dominate in disks during the
nonlinear evolution of the VSI, and current simulations must be
interpreted with care. 

\begin{figure}
\begin{center}
\includegraphics[scale=0.43]{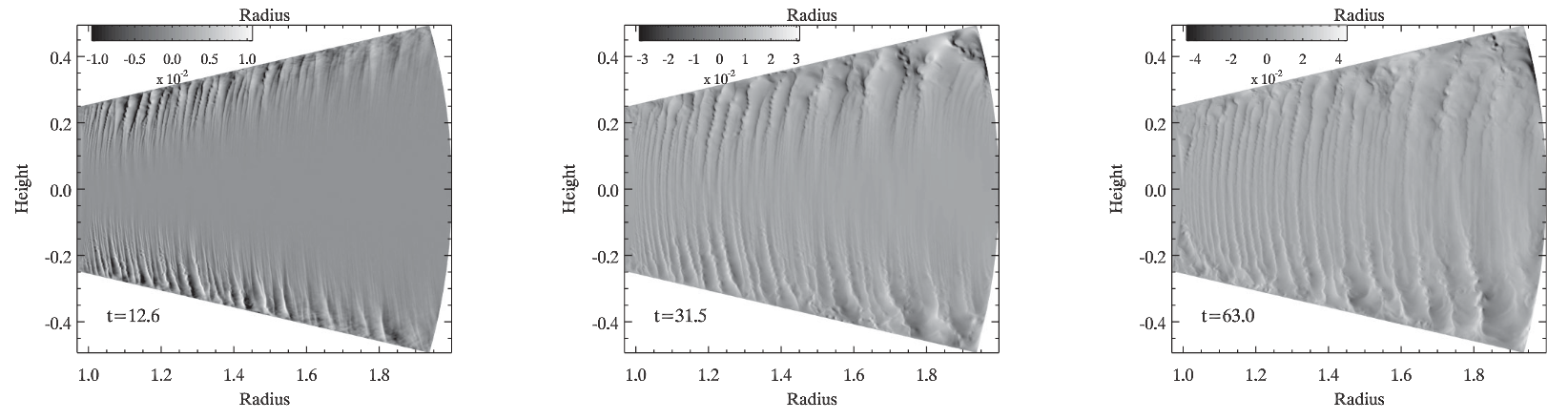}
\caption{Vertical velocity perturbations snapshots in the $(R,z)$
  plane in a global simulation of a locally isothermal disk during the
  growth of the vertical shear instability. Time increases from left to
  right. From \citet{nelsonetal13}.}
\label{fig:vsi_sims}
\end{center}
\end{figure}

Despite these concerns, a number of global numerical simulations have
been published that investigate the nonlinear outcome of the
VSI. \citet{nelsonetal13} showed that the VSI manifests itself as short
radial wavelengths perturbations (with comparatively larger vertical
scale) that first develop in the disk upper layers before eventually
affecting the entire disk (figure~\ref{fig:vsi_sims}). Turbulence ensues,
and lead to angular momentum transport rates that depends on the disk
cooling time. An upper bound of $\alpha \sim 10^{-3}$ is found in the
locally isothermal limit for which $N_z^2$ vanishes
\citep{nelsonetal13}, while values ranging from a few times $10^{-5}$
to a few times $10^{-4}$ are found for finite values of the cooling
time \citep{richardetal16}. In this situation, the VSI appears to
saturate via the triggering of short lived SBI-unstable vortices. This
is not unexpected, since the conditions of existence for the VSI and
the SBI are relatively similar, and suggest that the VSI could act as
a seed for the SBI. Finally, we note that more realistic simulations
that include heating by the central star and radiative cooling by dust 
particles are now feasible 
\citep{stoll&kley14,stoll&kley16}: the typical $\alpha \sim $ a few
times $10^{-4}$ that have been reported are in agreement with the
idealized simulations described above.

\section{Alternative routes to turbulence}
\label{sec:misc_proc}

At this stage, it should be clear that the complexity of accretion
disks is such that several physical processes have the potential to
affect their dynamics. The reader should thus not be
surprised to learn that several other possibilities have been proposed
in the literature that might lead to angular momentum transport in
disks. Some of them apply only in special situations (for
example, the gravitational instability requires that the
mass of the disk itself be significant), while other have only been
very recently proposed and must be investigated more thoroughly. In
the present section, we briefly review those of them that are the most
intensively discussed in the literature. We omit on purpose a
detailed discussion of the Rossby Wave Instability
\citep[RWI,][]{lovelaceetal99}. Although it may well be important for
some aspects of accretion disks dynamics and observations, the RWI is
confined to the vicinity of vortensity extrema and unlikely to have a
global impact on angular momentum transport in disks.

\paragraph{The gravitational instability (GI)}
\quad \\
So far, we restricted the discussion to massless accretion
disk. However, the disk self-gravity can render the flow unstable to a
gravitational instability (GI) in massive disks and could have some
implications for the dynamics of PP disks and for that of AGN
disks. The properties of GIs in disks have been much investigated, and
we point the interested reader to the recent reviews by
\citet{kratter&lodato16} for the case of PP disks, and by \citet{lodato12} 
for the case of AGN disks. Here, we briefly recall the main points, focusing
of those aspects of the problem that are relevant for angular momentum
transport. The importance of self-gravity is measured using the Toomre
parameter \citep{toomre64}: 
\begin{equation}
Q=\frac{c_s \kappa}{\pi G \Sigma} \sim \left( \frac{H}{R} \right)
\left( \frac{M}{M_d} \right) \, ,
\label{eq:toomre}
\end{equation} 
where $G$ is the gravitational constant, $c_s$ is the speed of sound
and $\Sigma$ is the disk surface density. Accretion disks are linearly
unstable to axisymmetric perturbations when $Q$ is smaller than unity
and linearly unstable to non-axisymmetric perturbations when $Q$ is of
order unity. The second relation in Eq.~(\ref{eq:toomre}) above, in
which $M_d$ stands for the disk mass contained within a radius R,
shows that this situation arises when the enclosed disk mass with a
radius $R$ is of order $(H/R)M$. For this reasons, GIs are expected to
develop preferentially in disks outer parts and are unlikely to
explain angular momentum transport in disks inner parts.  

While these linear arguments are fairly simple and well established,
understanding how GIs saturate is much more complicated. As shown by
Eq.~(\ref{eq:toomre}), $Q$ depends on the disk temperature through
$c_s$, so that the outcome of the instability depends once more on the
disk heating and cooling processes. For AGN disks that are not
irradiated, heating results from the thermalization of the kinetic
energy of the gas motions generated by the GI itself. The saturation
of the instability depends on the cooling timescale $\tau_c$ in the
sense that the disk reaches a quasi steady state in which heating and
cooling balance each other \citep{paczynski78}. The rate of angular
momentum transport in that case was numerically found to be
\citep{gammie01}:
\begin{equation}
\alpha \sim \frac{4}{9 \gamma (\gamma - 1) \Omega \tau_c} \, .
\end{equation}
This result is only marginally modified when irradiation is
important in the energy budget (as is the case for PP disks), as shown
by means of idealized 2D simulations by \citet{riceetal11}. When
$\tau_c=10 \Omega^{-1}$ (such as is typical of PP disks midplanes),
the above formula gives $\alpha \sim 10^{-2}$, thereby possibly
accounting for angular momentum transport 
in disks outer parts provided the disk mass is large enough. For
short cooling timescales, $\alpha$ can reach large values, and it
has been found early on in this case that disks can fragment and form
bound structures \citep{gammie01}. This has been
interpreted as a possible route to form planets in PP disks, 
although the issue is widely debated in the literature and depends on
subtle numerical issues \citep{kratter&lodato16}.

\paragraph{Vertical convection}
\quad \\
As mentioned in section~\ref{sec:hydro_stab}, accretion disk can
become convectively unstable to vertical convection if $N_z^2<0$ 
and this has historically been considered as a possible avenue toward
angular momentum transport \citep{vila78,lin&pap80}. While
early simulations suggested that angular momentum transport was
directed inward rather than outward
\citep{kleyetal93,stone&balbus96}, this surprising result was more 
recently shown to be due to the Rayleigh number $Ra$, which compares
the relative importance of buoyancy and dissipative processes, being
too small. At large Rayleigh number (such as found in accretion
disks), outward transport is recovered, with typical $\alpha$ values
of order a few times $10^{-4}$ \citep{lesur&ogilvie10}. All
these results, however, leave aside the question of the maintenance of
a negative vertical entropy gradient in the disk, which requires a
source of heating in the disk midplane region. 

Recently, \citet{bodoetal12} and \citet{hiroseetal14} both suggested,
by means of MHD simulations, that such a heating could be provided by
MRI turbulence. Although more work is needed to assess its
astrophysical implications, these results open the possibility for an
interesting coupling between the MRI and convection that
would affect the dynamics of erupting disks, such as in dwarf novae
and/or soft X-ray transients.

\paragraph{The Zombie vortex instability (ZVI)}
\quad \\
The ZVI was first identified in anelastic
simulations \citep{barrancoetal05}. This non-linear instability
appears in vertically stratified rotating sheared flows which are
stable for convection (i.e. $N_z^2>0$). The presence of vertical
buoyancy implies that in addition to epicyclic modes, buoyancy modes
are also present in the system. As shown by \cite{marcusetal13}, an
isolated vortex in such a flow spontaneously triggers non-axisymmetric
buoyancy waves which eventually break in critical
layers. \cite{marcusetal13} proposed that these critical layers were
in turn unstable and could generate new vortices, restarting the cycle
and filling the entire domain with vortices. 

This process was observed both in Boussinesq \citep{marcusetal13} and
compressible simulations \citep{marcusetal15}, and found to be
linearly associated with unstable buoyancy waves
\citep{umurhanetal16b}. However, critical 
layers being singularities of the linear inviscid equations, they are
necessarily regularised by viscosity and non-linear effects. As
expected, the ZVI is therefore highly sensitive to the diffusion
operator used in numerical simulations, and seems to be inexistent
when second order dissipation is used to model the flow. In addition,
thermal diffusion must be sufficiently small to avoid the damping of
buoyancy waves required by this instability
\citep{lesur&latter16}. Whether the ZVI is applicable to
PP disks is therefore a very open debate.

\paragraph{The stratorotational instability (SRI)}
\quad\\
The SRI occurs in situations similar to the
one of the ZVI: rotating shear flow with a stable vertical
stratification, except that it requires some sort of reflecting
boundary condition in the radial direction. It is a linear
instability, first identified in Taylor-Couette flows
\citep{YMM01}. The instability is primarily due to a resonance
between Kelvin waves which are traveling along the radial boundaries
(walls). For this resonance to be effective, the walls have to be
sufficiently close. The linear analysis shows that unstable modes have
a vertical wavelength $\lambda_z$ correlated to the wall separation: 
\begin{equation}
\lambda_z\sim \frac{\Omega}{N_z}L_\mathrm{wall}	,
\end{equation}
where $L_\mathrm{wall}$ is the radial distance between the walls
\citep{umurhan06}. Since PP disks always have $N_z\lesssim
\Omega$ \citep{dubrulleetal05b}, the stratorotational instability
can only grow when $L_\mathrm{wall}<H$. 

With an astrophysical application in mind, it is possible to
generalise the notion of rigid walls by assuming that a low density
parcel of gas is radially sandwiched between two parcels of high
density gas. In this case, the stratorotational instability still
shows up, but the required $L_\mathrm{wall}$ then decreases as the
density contrast decreases, leading to the suppression of the
instability when the parcels have equal densities
\citep[Fig. 68]{lesurPhD}.

Overall, these results suggest that the SRI might appear in those
regions of accretion disks where the flow is radially confined, such as
narrow gaps that are created by a planet, but that it is not a robust
way to transport angular momentum in the entire disk.

\paragraph{The Papaloizou-Pringle instability (PPI)}
\quad\\
Discovered by \citet{pap&pringle84}, the PPI relies on a
resonance between spiral density waves exchanging energy
accross the corotation radius of the disk
\citep{goldreichetal86}. Like the SRI, it depends on the presence of 
radial boundaries in the disk, so that it is equally unlikely to be a
robust mechanism for the radial transport of angular momentum in
disks. For more details on the physical mechanism of the PPI, we refer
the interested reader to the review article of \citet{pap&lin95}.

\section{Conclusion}

The length and diversity of this review shows that the research
community is not short of ideas to account for angular momentum
transport in Keplerian shear flows via turbulence, even within the
somewhat restrictive framework of hydrodynamics. Some of the proposed
solutions (like the subcritical transition - see
sec.~\ref{sec:subcrit_turb}) are heavily inspired from the traditional
field of hydrodynamics, while some others (vertical shear instability,
baroclinic instability, or gravitational instability) exploit some of
the peculiar properties of the astrophysical objects themselves. However,
at present time, it is fair to say that none of them (including the
MRI) appears to hold the promise of explaining - alone - how angular
momentum is transported outward in accretion disks. It is likely that
the later results from a combination of several of these processes
that operate simultaneously or at different locations in disks. For
example, in PP disks, the most recent results suggest to combine the
MRI, likely modified by the Hall term, and the VSI (which might excite 
the SBI, as discussed in sec~\ref{sec:vsi}), with GIs and/or infall
onto the disk \citep{lesuretal15} possibly playing a role in the outer
parts. In view of this complexity, direct observations of turbulence
in PP disks, now starting to be feasible with ALMA, will be of paramount
importance. Ironically, the first such observations suggest very low 
levels of turbulence \citep{flahertyetal15,pinteetal16}, incompatible
with any of the mechanisms discussed in this review as well as with our
current understanding of the MRI.

At the more fundamental level, the reason for the apparent
hydrodynamic stability of Keplerian flows remains elusive today. This 
is particularly surprising given the huge reservoir of free energy
that is available and must be understood. Progress will have to come
simultaneously from all sides of the battlefront, be it experimental,
numerical or theoretical, as each one suffer from its own
limitations. For example, experiments less sensitive to the end caps,
such as proposed by \citet{leclercqetal16}, must be
repeated at larger Reynolds numbers. They will have to be complemented
by better resolved direct numerical simulations that will allow, thanks
to ever increasing computational progress, higher $Re$ to be
explored. Theoretical work is also needed, and should take advantage,
as discussed in section~\ref{sec:subcrit_turb}, of the rapidly
expanding literature devoted to the question of shear flow stability
in the hydrodynamic community \citep[see for example the review of][on
the transition to turbulence in pipe flows]{eckhardtetal07}. Another
avenue which could be explored more systematically is the sensitivity
of Keplerian flows to external perturbations. Even if accretion
disks are nonlinearly stable, they might be subject to turbulence
triggered by low amplitude motions in the disk environment which are
subsequently amplified \citep{ioannou&kakouris01}. Only a 
combination of such widely different approaches will help settle the
question of whether or not Keplerian flows are vulnerable to a
subcritical transition to turbulence that Jean-Paul Zahn tried to
tackle $20$ years ago.

\section*{Acknowledgements}

The authors warmly acknowledge Adrian Barker \& Henrik Latter
for many discussions related to the physical processes presented in
this paper.

\bibliographystyle{astron}
\bibliography{author_hydro}

\end{document}